\documentclass[prd,amsmath,showkeys,showpacs,preprintnumbers,twocolumn,nofootinbib]{revtex4-1}

\usepackage{amssymb}
\usepackage{amsmath}
\usepackage{graphicx}
\usepackage{bm}

\newcommand{\INTM}{\int\frac{d^3\bm p}{(2\pi)^3}}

\begin{document}

\title{The Critical End Point of Quantum Chromodynamics Detected by Chirally Imbalanced
Quark Matter}
\author{Marco Ruggieri}
\affiliation{Yukawa Institute for Theoretical Physics, Kyoto
University, Kitashirakawa Oiwake-cho, Sakyo-ku, Kyoto 606-8502,
Japan.}

\begin{abstract}
We suggest the idea, supported by concrete calculations within
chiral models, that the critical endpoint of the phase diagram of
Quantum Chromodynamics with three colors can be detected, by means
of Lattice simulations of grand-canonical ensembles with a chiral
chemical potential, $\mu_5$, conjugated to chiral charge density.
In fact, we show that a continuation of the critical endpoint of
the phase diagram of Quantum Chromodynamics at finite chemical
potential, $\mu$, to a critical end point in the
temperature-chiral chemical potential plane, is possible. This
study paves the way of the mapping of the phases of Quantum
Chromodynamics at finite $\mu$, by means of the phases of a
fictitious theory in which $\mu$ is replaced by $\mu_5$.
\end{abstract}

\preprint{YITP-11-39}

\maketitle

\section{Introduction}
The critical endpoint, CP, of Quantum Chromodynamics
(QCD)~\cite{Asakawa:1989bq} is one of the most important aspects
of the phase diagram of strongly interacting matter. It is thus
not surprising that an intense experimental activity is nowadays
dedicated to the detection of such a point, which involves the
large facilities at RHIC and LHC; moreover, further experiments
are expected after the development of FAIR at GSI. Several
theoretical signatures of CP have been
suggested~\cite{Sig-CEP,Stephanov:1999zu}. Despite the importance
of CP, a firm theoretical evidence of its existence is still
missing. In fact, the sign problem makes the Lattice Monte Carlo
simulations difficult, if not impossible, in the large
baryon-chemical potential ($\mu$) region for $N_c
=3$~\cite{LQCD-CEP}, see~\cite{deForcrand:2010ys} for a recent
review. Therefore, it has not yet been possible to prove
unambiguously the existence and the location of CP starting from
first principles simulations of grand-canonical ensembles. The
strong coupling expansion of Lattice
QCD~\cite{SC-LQCD,deForcrand:2009dh} seems promising. Even more,
the predictions of effective models are spread in the $T-\mu$
plane, see for example~\cite{Stephanov:2007fk,Ohnishi:2011jv}.

An interesting overcoming of the sign problem for the quest of CP
is offered by analytic continuation of data obtained at imaginary
chemical potential, $\mu_I$~\cite{Alford:1998sd,de
Forcrand:2002ci,D'Elia:2002gd}. Recent promising analysis shows
that it might be possible to continue the critical line from the
region of imaginary $\mu$ to that of real
$\mu$~\cite{deForcrand:2008vr}, pinning down the critical point.
Another fruitful approach is given by simulations at finite
isospin chemical potential, $\delta\mu$, see for
example~\cite{Kogut:2002zg,deForcrand:2007uz,Cea:2009ba}. Even in
this case, a critical endpoint there appears before the transition
to the pion condensed phase. The latter point has been overlooked
and it has not yet been detected by mean field model calculations;
hence it certainly deserves further study. It is also worth to
cite the possibility to perform simulations in canonical, rather
then grand-canonical, ensembles. Preliminary results in this
direction have been presented recently in~\cite{Li:2011ee}. On the
purely theoretical side, it has been suggested very
recently~\cite{Hanada:2011ju} that the use of orbifold equivalence
in the large $N_c$ approximation of QCD can lead to relations
between the coordinates of CP at finite chemical potential, with
those at finite isospin chemical potential.

In this Article, we suggest a new, theoretical way to detect the
CP, by means of Lattice simulations with $N_c = 3$, which can be
considered as an alternative to the $\mu_I$ technique. In order to
accomplish this important program, we suggest to simulate QCD with
a chiral chemical potential, $\mu_5$, conjugated to the chiral
charge density, $n_5 = n_R - n_L$,
see~\cite{McLerran:1990de,Fukushima:2008xe,Fukushima:2010fe,Chernodub:2011fr}
for previous studies. Our idea, supported by concrete calculations
within microscopic effective models, is that CP can be {\em
continued} to a critical endpoint at $\mu_5 \neq 0$ and $\mu = 0$,
that we denote by CP$_5$, the latter being accessible to $N_c = 3$
Lattice QCD simulations of grand-canonical
ensembles~\cite{Fukushima:2008xe}. Therefore, the detection of the
former endpoint via Lattice simulations, can be considered as a
signal of the existence of the latter. To facilitate exposition,
we introduce the symbol ${\cal W}_5$ to denote the world with
$\mu=0,~\mu_5\neq0$. On the other hand, we will use the symbol
${\cal W}$ to denote the world with $\mu_5 = 0$, and which
corresponds to the physical universe.

The model calculations, in particular the ones based on the
Nambu-Jona-Lasinio model with the Polyakov
loop~\cite{Fukushima:2003fw} (PNJL model in the following) with
tree level coupling among chiral condensate and Polyakov
loop~\cite{Sakai:2010rp}, give numerical relations among the
coordinates of CP$_5$ and those of CP. In particular, the critical
temperature turns out to be almost unaffected by the process of
continuation; the critical value of the chemical potential,
$\mu_c$, on the other hand turns out to be almost half of the
critical chiral chemical potential, $\mu_{5c}$.

Before discussing our results, it is important to spend some word
more about the chiral chemical potential. In particular, we are
aware that ${\cal W}_5$ should be considered as a fictional
universe. As a matter of fact, $\mu_5$ cannot be considered as a
true chemical potential because, in the confinement phase, the
chiral condensate $\langle\bar qq\rangle$ mixes left- and
right-handed components of the quark field, leading to
non-conservation of $n_5$. This statement is true also in the
Quark-Gluon-Plasma phase, where the chiral condensate is much
smaller than its value in the confinement phase; in this case, the
non-conservation of $n_5$ is much softer, and mainly due to the
bare quark mass, $m \ll T$ with $T$ corresponding to the
temperature of the heath bath in which the fields live. Therefore,
the point of view that we adopt in this Article is to consider
$\mu_5$ as a mere mathematical artifice. However, ${\cal W}_5$
with $N_c = 3$ can be simulated on the Lattice. Therefore, for the
continuity property cited above, it is worth to study it by
grand-canonical ensemble simulations: it might furnish an evidence
of the existence of the critical endpoint in the real world.
Furthermore, once CP$_5$ is detected, it might be possible to make
use of Lattice simulations to detect inhomogeneous
phases~\cite{Nickel:2009ke,Flachi:2010yz} which could develop
around CP$_5$, as a continuation of those which develop at CP. The
results would then be of vital importance to understand, for
example, the inner structure of compact stellar objects. For the
aforementioned reasons, this study is very far from being of
purely academic or theoretical interest.

\section{Chiral Models}
Because of its non-perturbative nature, we cannot make first
principles calculations within QCD in the regimes to which we are
interested in, namely moderate $T$, $\mu$ and $\mu_5$. Hence we
need to rely on some effective model, which is built in order to
respect (at least some of) the symmetries of the QCD action. To
this end, we make use of the celebrated Quark-Meson
model~\cite{Gervais:1969zz}, and of the Nambu-Jona-Lasinio
model~\cite{Nambu:1961tp} (see~\cite{revNJL} for reviews) improved
with the Polyakov loop~\cite{Fukushima:2003fw}, dubbed PNJL model,
which have been used many times in recent years to describe
successfully the thermodynamics of QCD with two and two-plus-one
flavors,
see~\cite{Sakai:2010rp,Roessner:2006xn,Sasaki:2006ww,Abuki:2008nm,
Kashiwa:2007hw,Herbst:2010rf,Kahara:2008yg,Skokov:2010uh,Andersen:2011pr}
and references therein. They are interesting because they allow
for a self-consistent description of spontaneous chiral symmetry
breaking; even more, in the case the model is improved with the
Polyakov loop, it allows for a simultaneous computation of
quantities sensible to confinement and chiral symmetry breaking.

In this Section we describe both of the models. We firstly discuss
the Quark-Meson model that we use in our computation, which is
simpler to implement since take its simplest version, without the
complications due to the finite value of the quark masses and to
the Polyakov loop. This is an useful preparation to the more
important case of the PNJL model, which is more trustable
quantitatively since it is tuned to reproduce Lattice data at zero
and imaginary chemical potential.

\subsection{Quark-Meson Model}
The Quark-Meson (QM) model consists of the $O(4)$ linear sigma
model coupled to dynamical quarks via a Yukawa-type interaction.
The lagrangian density is given by
\begin{eqnarray}
{\cal L} &=& \bar q \left[i\partial_\mu\gamma^\mu - g(\sigma +
i\gamma_5\bm\tau\cdot\bm\pi) + \mu_5 \gamma^0\gamma^5 +
\mu\gamma^0\right] q
\nonumber \\
&& + \frac{1}{2}\left(\partial_\mu\sigma\right)^2 +
\frac{1}{2}\left(\partial_\mu\bm\pi\right)^2 - U(\sigma,\bm\pi)~.
\label{eq:LD1}
\end{eqnarray}
In the above equation, $q$ corresponds to a quark field in the
fundamental representation of color group $SU(3)$ and flavor group
$SU(2)$; besides, $\sigma$ and $\bm\pi$ correspond to the scalar
singlet and the pseudo-scalar iso-triplet fields, respectively. We
have a introduced chemical potential for the quark number density,
$\mu$, and a pseudo-chemical potential conjugated to chirality
imbalance, $\mu_5$.

We explain in some detail the physical meaning of the latter. The
quantity conjugated to $\mu_5$, namely the chiral charge density,
is given by $n_5 = n_R -n_L$, and represents the difference in
densities of the right- and left-handed quarks. At finite $\mu_5$,
a chirality imbalance is created, namely $n_5 \neq 0$. For
example, in the massless limit and at zero baryon chemical
potential one has~\cite{Fukushima:2008xe}
\begin{equation}
n_5 = \frac{\mu_5^3}{3\pi^2} + \frac{\mu_5 T^2}{3}~.
\end{equation}
If quark mass (bare or constituent) is taken into account, the
relation $n_5(\mu_5)$ cannot be found analytically in the general
case, and a numerical investigation is needed, see for
example~\cite{Fukushima:2010fe}. The imbalance of chiral density
can be created by instanton/sphaleron transition in QCD,
see~\cite{Fukushima:2008xe} and references therein. As a matter of
fact, non perturbative background gluon configurations with
nonzero winding number, $Q_W$, change the chirality of quarks
according to the Ward identity:
\begin{equation}
\frac{d n_5}{d t} = -\frac{g^2 N_f}{16\pi^2}\int d^3x F_{\mu\nu}^a
\tilde{F}_a^{\mu\nu} = - \frac{Q_W}{2 N_f}~.
\end{equation}
As a consequence, the addition of $\mu_5$ to the lagrangian
density of the chiral models mimics the instanton/sphaleron
induced chirality transitions.

The potential $U$ describes tree-level interactions among the
meson fields. In this Article, we restrict ourselves to the QM
model in the chiral limit for simplicity, and take
\begin{equation}
U(\sigma,\bm\pi) = \frac{\lambda}{4}\left(\sigma^2
+\bm\pi^2-v^2\right)^2~, \label{eq:U}
\end{equation}
which is invariant under the chiral group.

We work in the one-loop approximation, which amounts to consider
mesons as classical fields, and integrate only over fermions in
the generating functional of the theory to obtain the Quantum
Effective Potential (QEP). In the integration process, the meson
fields are fixed to their classical expectation values,
$\langle\bm\pi\rangle = 0$ and $\langle\sigma\rangle \neq 0$. The
physical value of $\langle\sigma\rangle$ will be then determined
by minimization of the QEP. The field $\sigma$ has the quantum
numbers of the QCD chiral condensate, $\langle\bar qq\rangle$.
Hence its non-vanishing expectation value breaks chiral symmetry
spontaneously, mimicking the chiral symmetry breaking of the QCD
vacuum.

The one-loop QEP in presence of $\mu_5$ has been discussed several
times~\cite{Fukushima:2008xe,Fukushima:2010fe,Chernodub:2011fr}.
The addition of the $\mu-$dependence is a textbook matter. The
final result is
\begin{eqnarray}
V &=& U -N_c N_f\sum_{s=\pm 1}\INTM \omega_s  \nonumber \\
&&-\frac{N_c N_f}{\beta}\sum_{s=\pm
1}\INTM\log\left(1+e^{-\beta(\omega_s -\mu)}\right) \nonumber \\
&&-\frac{N_c N_f}{\beta}\sum_{s=\pm
1}\INTM\log\left(1+e^{-\beta(\omega_s +\mu)}\right)~,
\label{eq:QEP}
\end{eqnarray}
where
\begin{equation}
\omega_s = \sqrt{(|\bm p| s -\mu_5)^2 + m_q^2}~, \label{eq:iii}
\end{equation}
corresponds to the pole of the quark propagator, and $m_q = g
\sigma$ is the constituent quark mass; finally, the index $s$
denotes the helicity projection.

In right hand side of the first line of Equation~\eqref{eq:QEP}
the momentum integral corresponds to the vacuum quark fluctuations
contribution to the QEP. It is divergent, and it gives a
contribution at $T=0$ and $\mu_5 \neq 0$. For our purposes, it is
enough to treat the model as an effective description of the
infrared regime of QCD. Therefore we treat the divergence
phenomenologically, introducing a momentum cutoff, $M$, in the
vacuum term. This is equivalent to introduce a momentum-dependent
quark mass, which is nonzero and constant for momenta lower than
the cutoff, and zero for larger values of momenta, thus realizing
a rough approximation of the effective (momentum-dependent and
ultraviolet suppressed) quark mass of full
QCD~\cite{Politzer:1976tv}.

The parameters of the model are tuned as in~\cite{Ohnishi:2011jv}.
They are chosen in order to satisfy the requirements that $\sigma
= f_\pi$ in the vacuum,
\begin{equation}
\left.\frac{\partial V}{\partial\sigma}\right|_{\sigma = f_\pi} =
0~, \label{eq:V1}
\end{equation}
and $m_\sigma = 700$ MeV, where
\begin{equation}
\left.\frac{\partial^2 V}{\partial\sigma^2}\right|_{\sigma =
f_\pi} = m_\sigma^2~, \label{eq:V2}
\end{equation}
Moreover, the constituent quark mass in the vacuum is fixed to the
value $m_q = 335$ MeV, which allows to fix the numerical value of
$g = m_q/f_\pi$ with $f_\pi = 92.4$ MeV. Finally, the ultraviolet
cutoff is taken $M = 600$ MeV. This procedure fixes $\lambda =
2.73$ and $v^2 = -(617.7$ MeV$)^2$.

\subsection{PNJL model}
In the PNJL model, quark propagation in the medium is described by
the following lagrangian density:
\begin{equation}
{\cal L} =\bar q\left(i\gamma^\mu D_\mu - m\right)q + {\cal
L}_I~;\label{eq:1ooo}
\end{equation}
here $q$ is the quark Dirac spinor in the fundamental
representation of the flavor $SU(2)$ and the color group;
$\bm\tau$ correspond to the Pauli matrices in flavor space. A sum
over color and flavor is understood. The covariant derivative
embeds the QCD coupling with the background gluon field which is
related to the Polyakov loop, see below. Furthermore, we have
defined
\begin{equation}
{\cal L}_I = G\left[\left(\bar qq\right)^2 + \left(i\bar
q\gamma_5\bm\tau q\right)^2\right]~.\label{eq:1}
\end{equation}

One advantage to use the PNJL model is that it has access to the
expectation value of the Polyakov loop, that we denote by $L$,
which is sensible to confinement or deconfinement properties of a
given phase. In order to compute $L$ we introduce a static,
homogeneous and Euclidean background temporal gluon field, $A_0 =
iA_4 = i \lambda_a A_4^a$, coupled minimally to the quarks via the
QCD covariant derivative~\cite{Fukushima:2003fw}. Then
\begin{equation}
L = \frac{1}{3}\text{Tr}_c\exp\left(i\beta\lambda_a A_4^a\right)~,
\end{equation}
where $\beta = 1/T$. In the Polyakov gauge, which is convenient
for this study, $A_0 = i\lambda_3 \phi + i \lambda_8 \phi^8$;
moreover, for simplicity we take $L = L^\dagger$ from the
beginning as in~\cite{Roessner:2006xn}, which implies $A_4^8 = 0$.
We have then
\begin{equation}
L = \frac{1+2\cos(\beta\phi)}{3}~.
\end{equation}
For our purpose, that is mainly the location of the critical
endpoint, we expect that the approximation $L = L^\dagger$ is
sufficient; as a matter of fact, we have verified that in this
case we reproduce, at finite $\mu$, the location of the critical
endpoint obtained in~\cite{Sakai:2010rp}, where the same
parametrization of the model is used, and where $L \neq L^\dagger$
from the beginning.

In our computation we follow the idea implemented
in~\cite{Sakai:2010rp}, which brings to a Polyakov-loop-dependent
coupling constant:
\begin{equation}
G = g\left[1 - \alpha_1 L L^\dagger -\alpha_2(L^3 +
(L^\dagger)^3)\right]~,\label{eq:Run}
\end{equation}
The ansatz in the above equation was inspired
by~\cite{Kondo:2010ts,Frasca:2008zp} in which it was shown
explicitly that the NJL vertex can be derived in the infrared
limit of QCD, it has a non-local structure, and it acquires a
non-trivial dependence on the phase of the Polyakov loop. We refer
to~\cite{Sakai:2010rp} for a more detailed discussion. This idea
has been analyzed recently in~\cite{Braun:2011fw}, where the
effect of the confinement order parameter on the four-fermion
interactions and their renormalization-group fixed-point structure
has been investigated. The numerical values of $\alpha_1$ and
$\alpha_2$ have been fixed in~\cite{Sakai:2010rp} by a best fit of
the available Lattice data at zero and imaginary chemical
potential of Ref.~\cite{D'Elia:2009qz,Bonati:2010gi}. In
particular, the fitted data are the critical temperature at zero
chemical potential, and the dependence of the Roberge-Weiss
endpoint on the bare quark mass. The best fit procedure leads to
$\alpha_1 = \alpha_2 \equiv \alpha = 0.2 \pm 0.05$, within the
hard cutoff regularization scheme, which is the same scheme that
we adopt in this Article.

In the one-loop approximation, the effective potential of this
model is given by
\begin{eqnarray}
V &=& {\cal U}(L,L^\dagger,T) +\frac{\sigma^2}{G}  -N_c N_f\sum_{s=\pm 1}\INTM \omega_s \nonumber \\
&&-\frac{N_c N_f}{\beta}\sum_{s=\pm 1}\INTM\log\left(F_+ F_-\right) \nonumber \\
&&~\label{eq:OB}
\end{eqnarray}
where
\begin{eqnarray}
F_- &=& 1+3L e^{-\beta(\omega_s - \mu)} +3L^\dagger
e^{-2\beta(\omega_s - \mu)} + e^{-3\beta(\omega_s -
\mu)}~,\nonumber \\
&& \\
F_+ &=& 1+3L^\dagger e^{-\beta(\omega_s + \mu)} +3L
e^{-2\beta(\omega_s + \mu)} + e^{-3\beta(\omega_s +
\mu)}~,\nonumber \\
&&
\end{eqnarray}
denote the statistical confining thermal contributions to the
effective potential; $\omega_s$ is still given by
Equation~\eqref{eq:iii}, with $m_q = m -2\sigma$. Once again the
vacuum fluctuation term is regularized by means of a ultraviolet
cutoff, that we denote by $M$. The relation between the chiral
condensate and $\sigma$ in the PNJL model is $\sigma =
2G\langle\bar qq\rangle$.

We notice that in this case we take quarks with a finite bare
mass, which will be fixed by requiring that the pion mass in the
vacuum is in agreement with its experimental value. We also notice
that the PNJL model considered here, which is dubbed Extended-PNJL
in~\cite{Sakai:2010rp}, has been tuned in order to reproduce
quantitatively the Lattice QCD thermodynamics at zero and
imaginary quark chemical potential. Hence, it represents a
faithful description of QCD, in terms of collective degrees of
freedom related to chiral symmetry breaking and deconfinement.

The potential term $\mathcal{U}$ in Eq.~\eqref{eq:OB} is built by
hand in order to reproduce the pure gluonic lattice data with $N_c
= 3$~\cite{Roessner:2006xn}. We adopt the following logarithmic
form,
\begin{equation}
 \begin{split}
 & \mathcal{U}[L,\bar L,T] = T^4\biggl\{-\frac{a(T)}{2}
  \bar L L \\
 &\qquad + b(T)\ln\bigl[ 1-6\bar LL + 4(\bar L^3 + L^3)
  -3(\bar LL)^2 \bigr] \biggr\} \;,
 \end{split}
\label{eq:Poly}
\end{equation}
with three model parameters (one of four is constrained by the
Stefan-Boltzmann limit),
\begin{equation}
 \begin{split}
 a(T) &= a_0 + a_1 \left(\frac{T_0}{T}\right)
 + a_2 \left(\frac{T_0}{T}\right)^2 , \\
 b(T) &= b_3\left(\frac{T_0}{T}\right)^3 \;.
 \end{split}
\label{eq:lp}
\end{equation}
The standard choice of the parameters reads $a_0 = 3.51$, $a_1 =
-2.47$, $a_2 = 15.2$ and $b_3 = -1.75$. The parameter $T_0$ in
Eq.~\eqref{eq:Poly} sets the deconfinement scale in the pure gauge
theory. In absence of dynamical fermions one has $T_0 = 270$
\text{MeV}. However, dynamical fermions induce a dependence of
this parameter on the number of active
flavors~\cite{Herbst:2010rf}. For the case of two light flavors to
which we are interested here, we take $T_0 = 190$ MeV as
in~\cite{Sakai:2010rp}. Also for the remaining parameters we
follow~\cite{Sakai:2010rp} and take $M = 631.5$ MeV, $m=5.5$ MeV
and $G=5.498\times 10^{-6}$ MeV$^{-2}$.

\section{Critical endpoint at zero chemical potential}

\begin{figure}[t!]
\begin{center}
\includegraphics[width=8.5cm]{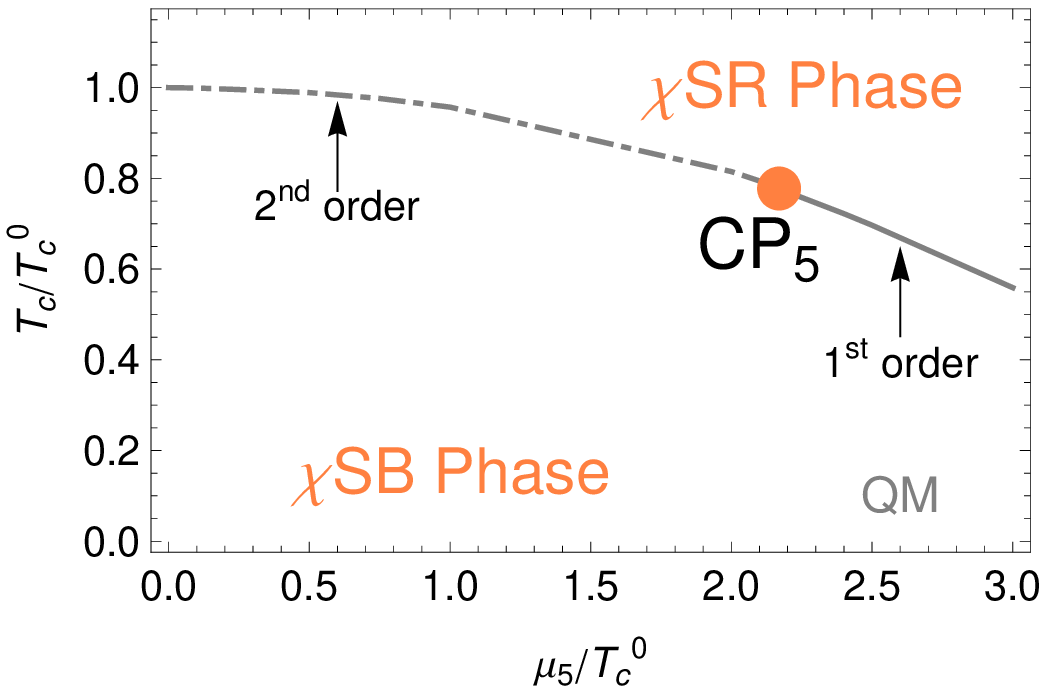}\\
\includegraphics[width=8.5cm]{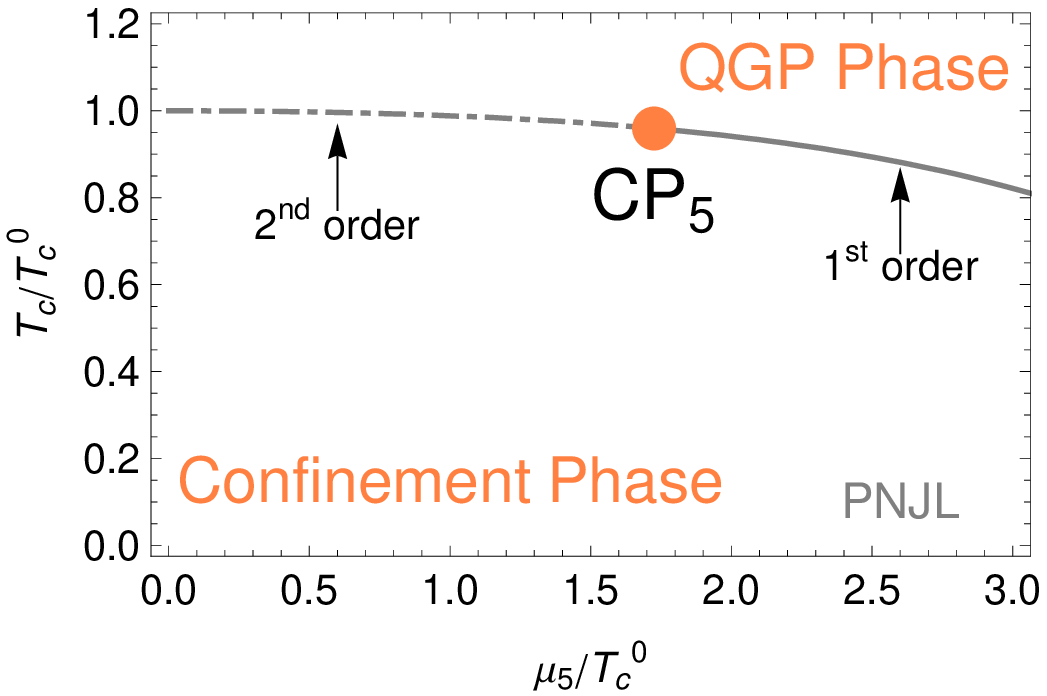}
\caption{\label{Fig:PD} ({\em Color online}). {\em Upper panel:}
Phase diagram of the QM model in the $\mu_5-T$ plane. The
dot-dashed line corresponds to the second order chiral phase
transition; the solid line denotes the first order phase
transition. The orange dot is the critical endpoint. The scale
$T_c^0 = 174.1$ MeV corresponds to the critical temperature at
$\mu_5 = 0$. {\em Lower panel:} Phase diagram of the PNJL model.
Lines convention is the same as in the upper panel. The scale
$T_c^0 = 173.9$ MeV corresponds to the critical temperature at
$\mu_5 = 0$. }
\end{center}\end{figure}

In Figure~\ref{Fig:PD} we plot the phase diagram of the chiral
models in the $\mu_5 - T$ plane, for the case $\mu = 0$. It is
obtained by a minimization procedure of the full
potential~\eqref{eq:QEP} for the QM model, and~\eqref{eq:OB} for
the case of the PNJL model.

In the case of the QM model, the critical temperature $T_c$ is
identified with the one at which $\langle\sigma\rangle_{T=T_c} =
0$. For the PNJL model, since chiral symmetry is broken explicitly
by the quark mass and the phase transitions are replaced by
crossovers, we identify the critical temperature with that at
which $dL/dT$ is maximum. We have checked that the latter deviates
from that at which $|d\sigma/dT|$ is maximum only of a few MeV, in
the whole range of parameters studied. Even in this case, with an
abuse of nomenclature, we dub the pseudo-critical lines as second
order and first order, as in the case of the QM model. It is clear
from the context that, whenever we talk about the PNJL model, the
term second order transition has to be taken as a synonym of
smooth crossover; similarly, the term first order transition is a
synonym of discontinuous jump of the order parameters.

In the case of the QM model, the data about the condensates as a
function of temperature at $\mu_5 \neq 0$ have been presented
already in the literature~\cite{Chernodub:2011fr}, therefore their
reproduction in this Article would be just of academic interest;
hence we skip this step here. Our focus is the discussion of the
critical endpoint and of its evolution from ${\cal W}_5$ to ${\cal
W}$; hence we focus on CP$_5$ and CP.

In both panels of Figure~\ref{Fig:PD}, the grey dashed line
corresponds to the second order chiral phase transition in the
case of the QM model, or to a smooth crossover in the case of the
PNJL model. The solid line, on the other hand, denotes the first
order transition. The dot corresponds to CP$_5$. In the following,
we label the coordinates of CP$_5$ by $(\mu_{5c},T_{5c})$. For the
case of the QM model, since we do not have any information about
confining-deconfining property of a given phase, we can label the
phases of the model only in terms of the chiral symmetry. We call
the phase below the critical line as the chiral symmetry broken
phase; similarly, above the critical line, chiral symmetry is
restored, hence we call this phase as the chiral symmetry restored
phase. It is then natural that CP$_5$ in this case is a {\it
chiral} critical endpoint.

\begin{figure}[t!]
\begin{center}
\includegraphics[width=8.5cm]{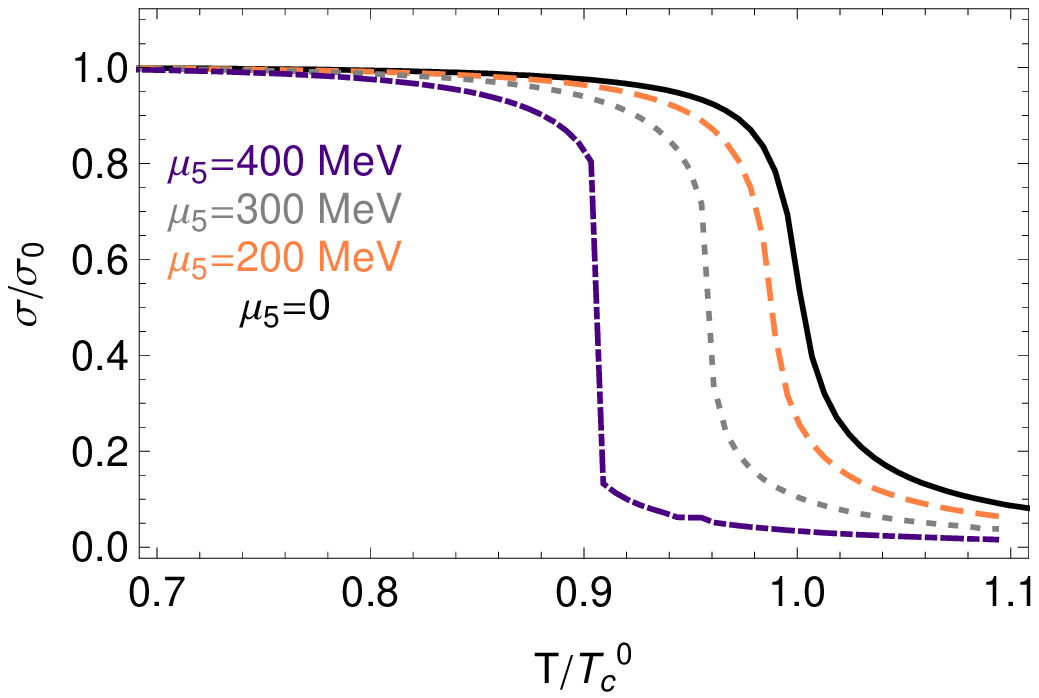}\\
\vspace{0.1cm}
\includegraphics[width=8.5cm]{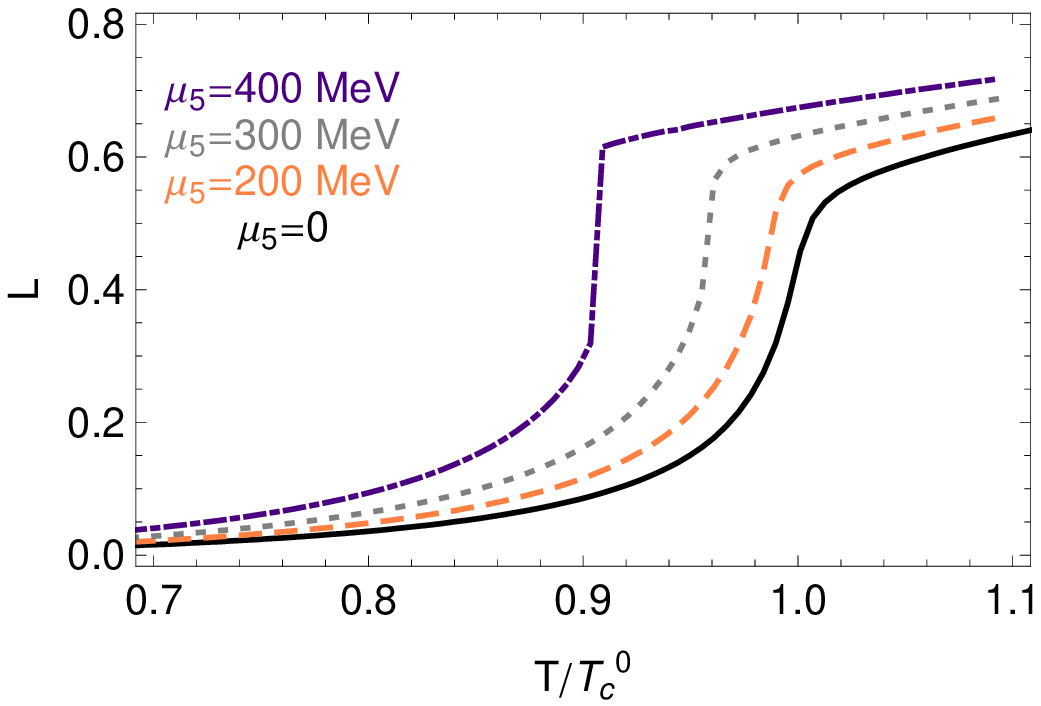}
\caption{\label{Fig:giaguara} ({\em Color online}). {\em Upper
panel:} Chiral condensate in the PNJL model, normalized to the
value at zero temperature, as a function of temperature for
several values of $\mu_5$. Black solid line corresponds to the
case $\mu_5 = 0$; orange dashed line to $\mu_5 = 200$ MeV; grey
dotted line to $\mu_5 = 300$ MeV; indigo dot-dashed line to $\mu_5
= 400$ MeV. {\em Lower panel:} Expectation value of the Polyakov
loop as a function of temperature, for several values of $\mu_5$.
Dashing and color convention is the same as in the upper panel.}
\end{center}\end{figure}

For the case of the PNJL model, on the other hand, we have access
to the chiral condensate and to the Polyakov loop expectation
value. As a consequence, we can label the phases of the model in
terms both of confining properties, and of chiral symmetry. In the
model at hand, because of the entanglement in
Equation~\eqref{eq:Run}, the deconfinement and chiral symmetry
restoration crossovers take place simultaneously both at value of
$\mu_5$. This is proved by our data about $\sigma$ and $L$, see
Figure~\ref{Fig:giaguara} in which we plot the chiral condensate
(upper panel) and the expectation value of the Polyakov loop
(lower panel) as a function of temperature, for several values of
$\mu_5$. In~\cite{Fukushima:2008xe} the two crossovers joined only
for $\mu_5 > \mu_{5c}$. Our results are a natural consequence of
the entanglement vertex in Equation~\eqref{eq:Run}, which was
neglected in~\cite{Fukushima:2010fe}.

It is useful to stress that in the PNJL model we are discussing
crossovers. Due to the crossover nature of the phenomena, a unique
definition of the critical temperatures is not available; in this
article we have used the definition which is easiest to implement,
namely the study of the peaks of the derivatives of the order
parameters. We dub these quantities as effective susceptibilities.
Within numerical error, we find that the peaks for the effective
susceptibilities of the Polyakov loop and of the chiral condensate
coincide in temperature, within few MeV. We cannot exclude that
using different definitions for the pseudo-critical temperatures,
like the identification of the crossovers with the peaks of true
susceptibilities, the former can be shifted of some MeV, leading
eventually to a larger split of the deconfinement and the
crossover temperatures. However, the qualitative picture should
not be modified drastically, as previous studies within the PNJL
model have shown.

Because of this peculiarity of the PNJL model, at the
pseudo-critical line both deconfinement and chiral restoration
crossovers take place. Hence the region below the pseudo-critical
line is characterized by confinement and spontaneous breaking of
chiral symmetry; we label this phase as the confinement phase. On
the other hand, the phase above the critical line is identified
with the Quark-Gluon-Plasma phase. In this case, CP$_5$ is both
{\it chiral} and {\it deconfinement} critical endpoint.

For what concerns the coordinates of CP$_5$, in the case of the QM
model we find
\begin{equation}
\left(\frac{\mu_{5c}}{T_c^0},\frac{T_c}{T_c^0}\right) =
(2.16,0.78)~,~~~~~\text{CP}_5~\text{(QM)}~, \label{eq:CP5qm}
\end{equation}
where $T_c^0 = 174.1$ MeV is the chiral symmetry restoration
temperature at $\mu = \mu_5 = 0$. Moreover, for the PNJL model we
find
\begin{equation}
\left(\frac{\mu_{5c}}{T_c^0},\frac{T_c}{T_c^0}\right) =
(1.73,0.96)~,~~~~~\text{CP}_5~\text{(PNJL)}~, \label{eq:CP5}
\end{equation}
where $T_c^0 = 173.9$ MeV is the deconfinement temperature at $\mu
= \mu_5 = 0$.

\section{Critical endpoint at finite chemical potential}
Next we turn to discuss the more general case with both $\mu_5$
and $\mu$ different from zero. Our scope is to show that, at least
within the models, CP naturally evolves into CP$_5$. Hence the
latter, if detected on the Lattice, can be considered as the
benchmark of the former. In particular the PNJL model, which is in
quantitative agreement with the Lattice at zero chemical
potential, gives a numerical relation among the coordinates of CP
and CP$_5$, which might be taken as a guide to estimate the
coordinates of CP in QCD, once CP$_5$ is detected.

\begin{figure}[t!]
\begin{center}
\includegraphics[width=8.5cm]{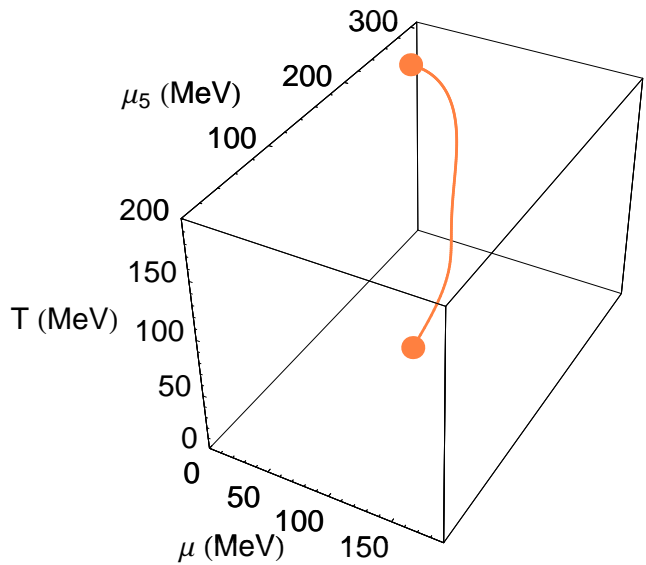}
\caption{\label{Fig:EV} ({\em Color online}). Evolution of the
critical endpoint in the $\mu-\mu_5-T$ space, for the PNJL model.}
\end{center}\end{figure}

\begin{figure}[t!]
\begin{center}
\includegraphics[width=8.5cm]{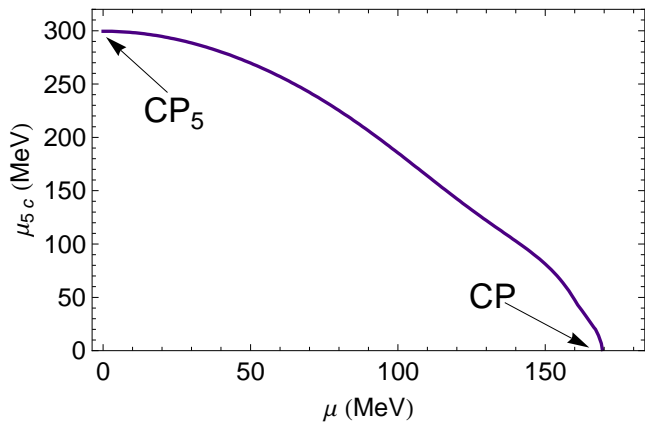}
\caption{\label{Fig:Proje} ({\em Color online}). Projection of
Fig.~\ref{Fig:EV} onto the $\mu - \mu_{5}$ plane, for the PNJL
model. The solid line corresponds to the $\mu_5$-coordinate of the
critical endpoint.}
\end{center}\end{figure}

In Figure~\ref{Fig:EV} we collect our data on the critical point
of the phase diagram in the $\mu-\mu_5-T$ space, in the case of
the PNJL model (for the QM model we obtain similar results). The
orange solid line is the union of the critical points computed
self-consistently at several values of $\mu$: at any value of
$\mu$, a point on the line corresponds to the critical point of
the phase diagram in the $\mu_5-T$ plane. Thus the line
pictorially describes the evolution of the critical point of the
chiral model at hand, from CP to CP$_5$. In Figure~\ref{Fig:Proje}
we plot a projection of Fig.~\ref{Fig:EV} onto the $\mu - \mu_{5}$
plane, for the PNJL model. The indigo solid line corresponds to
the $\mu_5$-coordinate of the critical endpoint. The critical
temperature is not so much affected when we continue CP$_5$ to CP
(we measure a change approximately equal to the $3\%$), therefore
the projection in the $\mu-T$ plane is redundant.

At $\mu_5 = 0$, the critical point is found at the following
coordinates:
\begin{equation}
\left(\frac{\mu_c}{T_c^0},\frac{T_c}{T_c^0}\right) =
(0.92,0.93)~,~~~~~\text{CP}~\text{(PNJL)}~, \label{eq:CP}
\end{equation}
which correspond to $\mu_c \approx 160$ MeV and $T_c \approx 165$
MeV, in agreement with the results of~\cite{Sakai:2010rp}.
Similarly for the case of the QM model we find
\begin{equation}
\left(\frac{\mu_c}{T_c^0},\frac{T_c}{T_c^0}\right) =
(1.58,0.49)~,~~~~~\text{CP}~\text{(QM)}~. \label{eq:CPqm}
\end{equation}

The natural question which arises is: suppose a grand-canonical
ensemble simulation finds CP$_5$; then, how can the coordinates of
the latter point help to locate CP? We can answer to this
important question within the PNJL model. Our numerical results,
Equations~\eqref{eq:CP5} and~\eqref{eq:CP}, suggest the following
relations:
\begin{equation}
\frac{\mu_c}{\mu_{5c}} \approx 0.53~,~~~\frac{T_c}{T_{5c}} \approx
0.97~,~~~~~~\text{(PNJL)}~. \label{eq:BBB}
\end{equation}
Within the QM model we obtain similar relations, but in our
opinion, those in~\eqref{eq:BBB} are more trustable quantitatively
because the PNJL model has been tuned to be in quantitative
agreement with Lattice data at zero as well as imaginary chemical
potential~\cite{Sakai:2010rp}, a characteristic which is not
satisfied by the QM model used here.

The model predictions~\eqref{eq:BBB} relate the coordinates of CP
to those of CP$_5$. In particular, it is interesting that the
critical temperature is almost unchanged in the continuation of CP
to CP$_5$. Of course, since these results are deduced by a model,
it is extremely interesting and important to study how
Equation~\eqref{eq:BBB} is affected by varying parameters like the
bare quark mass, or the number of active flavors. This observation
opens the possibility to develop further non-academic model
studies of the problem that we discuss in this Article.

\section{Discussion}
In this Section we summarize briefly the results obtained, and
discuss their conceptual relevance, and potential applications to
the Lattice as well.

Our main goal is to show that the continuation of the critical
endpoint of the QCD phase diagram, CP, to a fictitious critical
endpoint, CP$_5$, belonging to a phase diagram in the $\mu_5 - T$
plane, is reasonable. Here $\mu_5$ corresponds to the chiral
chemical potential, which is conjugated to the chiral density
imbalance, $n_5 = n_R - n_L$. As we have already stressed,
strictly speaking $\mu_5$ should be regarded as a pseudo-chemical
potential: quark condensate in the confinement phase induces a mix
among left- and right-handed quark field components. As a
consequence, the conjugated quantity to $\mu_5$, namely the chiral
charge density, is not conserved. Hence, it is legitimate to
consider $\mu_5$ as a mere mathematical artifact, and the world in
which $\mu$ is replaced by $\mu_5$, that we have baptized ${\cal
W}_5$, as an artificial universe, not necessarily related to the
physical world. This is the point of view that we adopt in this
Article.

However, even accepting this minimalist point of view, ${\cal
W}_5$ has the quality that it can be simulated on a Lattice.
Indeed, the sign problem which affects simulations of three color
QCD at finite chemical potential, does not affect QCD in ${\cal
W}_5$~\cite{Fukushima:2008xe}. Thus, it is possible to check
wether CP$_5$ there exists or not, within first principles
calculations. If Lattice simulations find CP$_5$, our result
suggests the simplest interpretation: CP$_5$ {\it would be nothing
but the continuation of the critical point which there exists in
QCD}. Hence we suggest to interpret CP$_5$, if found, as a signal
of the existence of the critical point in QCD. The coordinates of
CP are related to those of CP$_5$ by Equation~\eqref{eq:BBB}.

Furthermore, it is extremely intriguing, theoretically speaking,
that the critical endpoint CP$_5$ that we find in this study,
which confirms the findings of previous studies obtained within
different models, is the analytic continuation of the critical
endpoint, CP, of the theory at finite quark chemical potential.

As anticipated, previous
studies~\cite{Fukushima:2010fe,Chernodub:2011fr} have already
discussed the possible appearance of CP$_5$. In particular,
in~\cite{Fukushima:2010fe} a PNJL model in a strong magnetic
background, without entanglement vertex [see
Equation~\eqref{eq:Run}] has been considered. In that context, the
chiral chemical potential was introduced to mimic the presence of
chirality imbalance induced by instantons and sphalerons
transitions in the hot QCD medium. Quantitatively, our results on
the critical endpoint are in agreement with those
of~\cite{Fukushima:2010fe}.

In~\cite{Chernodub:2011fr}, the phase diagram in the $\mu_5 - T$
plane has been computed within the QM model improved with the
Polyakov loop. The models considered here and
in~\cite{Chernodub:2011fr} are different: the zero point energy is
not taken into account in~\cite{Chernodub:2011fr}, and on the
other hand we do not include the Polyakov loop in our QM model
calculations. For these reasons, it is not necessary to have a
quantitative agreement among our results and those
of~\cite{Chernodub:2011fr}. On the contrary, it is important to
notice that the qualitative picture is the same, namely the
existence of a critical endpoint in the $\mu_5 - T$ plane.

In both of the aforementioned studies, the quark chemical
potential $\mu$ has not been taken into account. Therefore, the
possibility to continue CP$_5$ to CP, which is the main idea of
the study illustrated in this Article, was not considered in those
references.

\section{Conclusions and Outlook}
In this Article, we have suggested the possibility of continuation
of the critical endpoint of the phase diagram of $N_c = 3$ QCD,
CP, to a critical endpoint dubbed CP$_5$ belonging to a fictional
world, ${\cal W}_5$, in which the quark number chemical potential
is replaced by a pseudo-chemical potential, $\mu_5$ conjugated to
the chiral charge density, $n_5$. The universe ${\cal W}_5$ has
the merit that it can be simulated on the
Lattice~\cite{Fukushima:2008xe} for $N_c = 3$. This suggestion is
based on concrete calculations within chiral models. In
particular, we have used the PNJL model with entanglement vertex,
introduced in~\cite{Sakai:2010rp}, which offers a description of
the QCD thermodynamics in terms of collective degrees of freedom,
which is in quantitative agreement with Lattice data at zero and
imaginary chemical potential.

Our main idea is that simulations in ${\cal W}_5$ might reveal the
existence of a critical endpoint, CP$_5$, in the phase diagram.
Then, this critical point might be interpreted as the continuation
of the critical point which is expected to belong to the phase
diagram of real QCD, because of the continuity summarized in
Fig.~\ref{Fig:EV}.  Hence it would be an indirect evidence of the
existence of the critical point in real QCD. Moreover, numerical
predictions of the model, which connect CP to CP$_5$, are in
Equation~\eqref{eq:BBB}. For these reasons, the interest of the
present study is very far from being only academic or purely
theoretical. Our result paves the way of the mapping of the phases
of Quantum Chromodynamics at finite $\mu$, by virtue of the phases
of a fictitious theory in which $\mu$ is replaced by $\mu_5$.

In our calculations there are some factors that we have not
included for simplicity, and that might be interesting to include
in more complete calculations (massive quarks, vector
interactions, just to cite a couple of examples). In view of a
possible mapping of the phase diagram of QCD using simulations of
grand-canonical ensembles in ${\cal W}_5$, it is of great interest
to extend the analysis of~\cite{Nickel:2009ke,Flachi:2010yz} about
inhomogeneous condensates, to CP$_5$. Moreover, in our opinion it
is important to understand quantitatively how the numerical
predictions of the model, namely Equation~\eqref{eq:BBB}, are
affected by varying the quark masses, and introducing a third
flavor. We plan to report on these topics in the next future.

{\bf Acknowledgements}. Part of this work was inspired by
stimulating discussions with H.~Warringa, who is acknowledged.
Moreover, we acknowledge M.~Chernodub, M.~D'Elia, P.~de Forcrand,
R.~Gatto, A.~Ohnishi, A.~Yamamoto and N.~Yamamoto for
correspondence, careful reading of the manuscript, criticism and
encouragement. This work is supported by the Japan Society for the
Promotion of Science under contract number P09028.


\end{document}